%% file: pns17_ampires.tex
\documentclass[a4paper]{jpconf}
\newcommand{\ros}{ROSAT}

\newcommand{\eROS}{eROSITA}

\def \msev{M7}
\usepackage{graphicx}
\usepackage{amsmath,amssymb}
\usepackage[square,sort&compress,numbers]{natbib}
\begin{document}
\nocite{*}
\title{The missing links of neutron star evolution in the \eROS\ all-sky X-ray survey}
\author{A~M~Pires} 
\address{Leibniz-Institut f\"ur Astrophysik Potsdam (AIP), An der Sternwarte 16, 14482 Potsdam, Germany}
\ead{apires@aip.de}
\begin{abstract}
The observational manifestation of a neutron star is strongly connected with the properties of its magnetic field. During the star's lifetime, the field strength and its changes dominate the thermo-rotational evolution and the source phenomenology across the electromagnetic spectrum.
Signatures of magnetic field evolution are best traced among elusive groups of X-ray emitting isolated neutron stars (INSs), which are mostly quiet in the radio and $\gamma$-ray wavelengths.
It is thus important to investigate and survey INSs in X-rays in the hope of discovering peculiar sources and the long-sought missing links that will help us to advance our understanding of neutron star evolution. The Extended R\"ontgen Survey with an Imaging Telescope Array (\eROS), the primary instrument on the forthcoming Spectrum-RG mission, will scan the X-ray sky with unprecedented sensitivity and resolution. The survey has thus the unique potential to unveil the X-ray faint end of the neutron star population and probe sources that cannot be assessed by standard pulsar surveys.
\end{abstract}
\section{Introduction}
An isolated neutron star (INS) radiates at the expense of its rotational, thermal, and magnetic energy. The emission is produced through various mechanisms, with photon energies covering the entire electromagnetic spectrum. Fundamental properties inherited at birth (mass, composition, spin rate, magnetic field, etc), as well as their subsequent temporal evolution, determine the dominant emission mechanism at a given age. These, however, are subject to many theoretical uncertainties and remain poorly constrained by observations.

The population of over 2,600 neutron stars observed in our Galaxy is dominated by radio pulsars \cite{man05}. Of those, about 50 peculiar X-ray emitting sources are not detected by radio and $\gamma$-ray pulsar surveys: they are the young and energetic magnetars, usually identified by their remarkable spectral properties and bursts of high energy emission \cite{kas17}; the local group of thermally emitting middle-aged INSs discovered by \ros\ and dubbed the magnificent seven (\msev) \citep{tur09}; and the young and seemingly weakly-magnetised central compact objects (CCOs), which are thermal X-ray sources located near the centre of supernova remnants \cite{got13}. These groups are important as they provide a privileged view of a variety of emission processes and evolutionary channels that cannot be probed through the bulk of the pulsar population. 
\begin{figure}[t]
\begin{minipage}{0.5\textwidth}
\includegraphics*[width=0.9\textwidth]{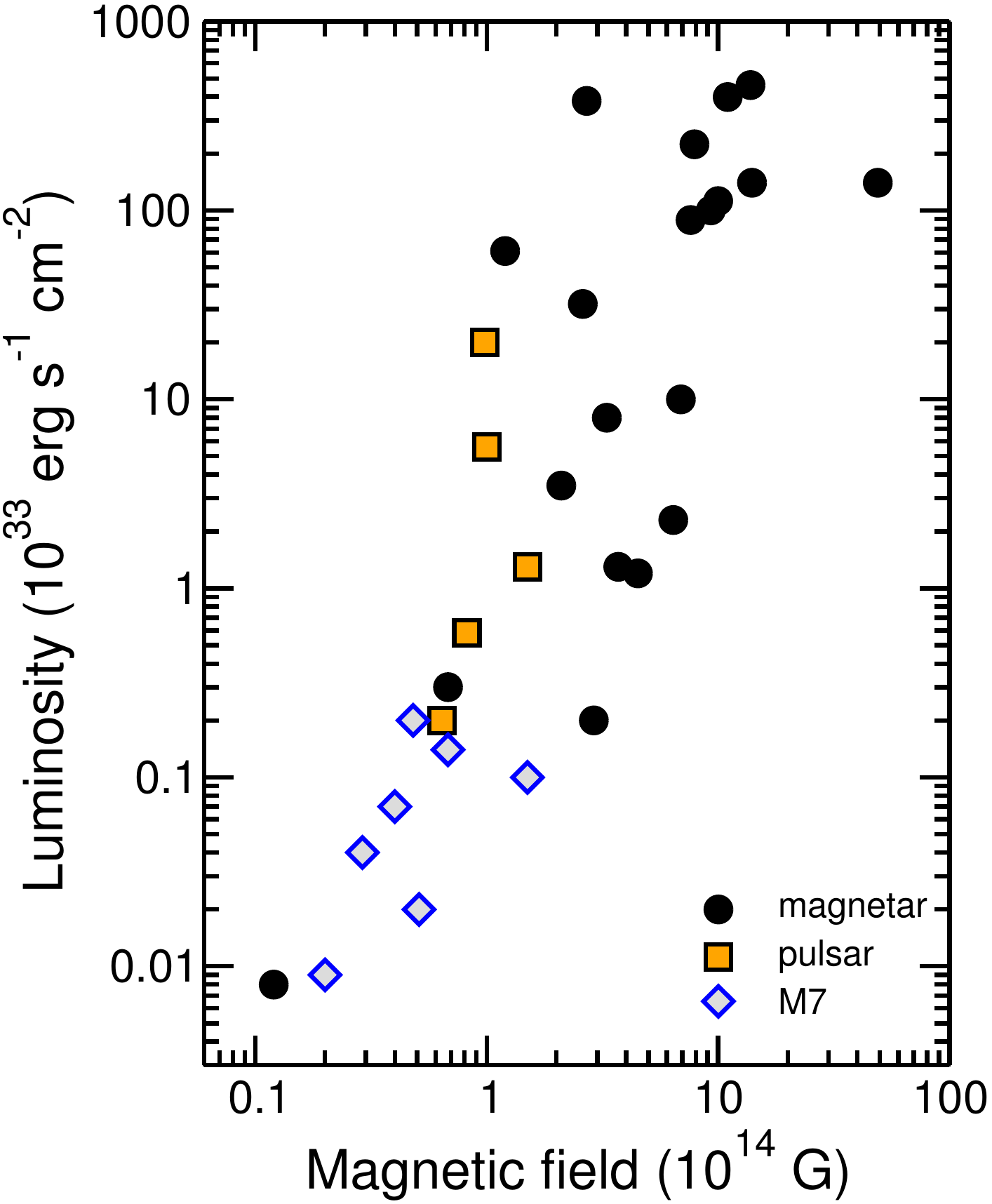}
\caption{\small Correlation between thermal luminosity and magnetic field intensity for high magnetic field INSs (data extracted from the Magnetar Outburst Online Catalog, \texttt{http://magnetars.ice.csic.es}). 
\label{fig_BLx}}
\end{minipage}
\hspace{0.05\textwidth}
\begin{minipage}{0.45\textwidth}
\includegraphics*[width=0.9\textwidth]{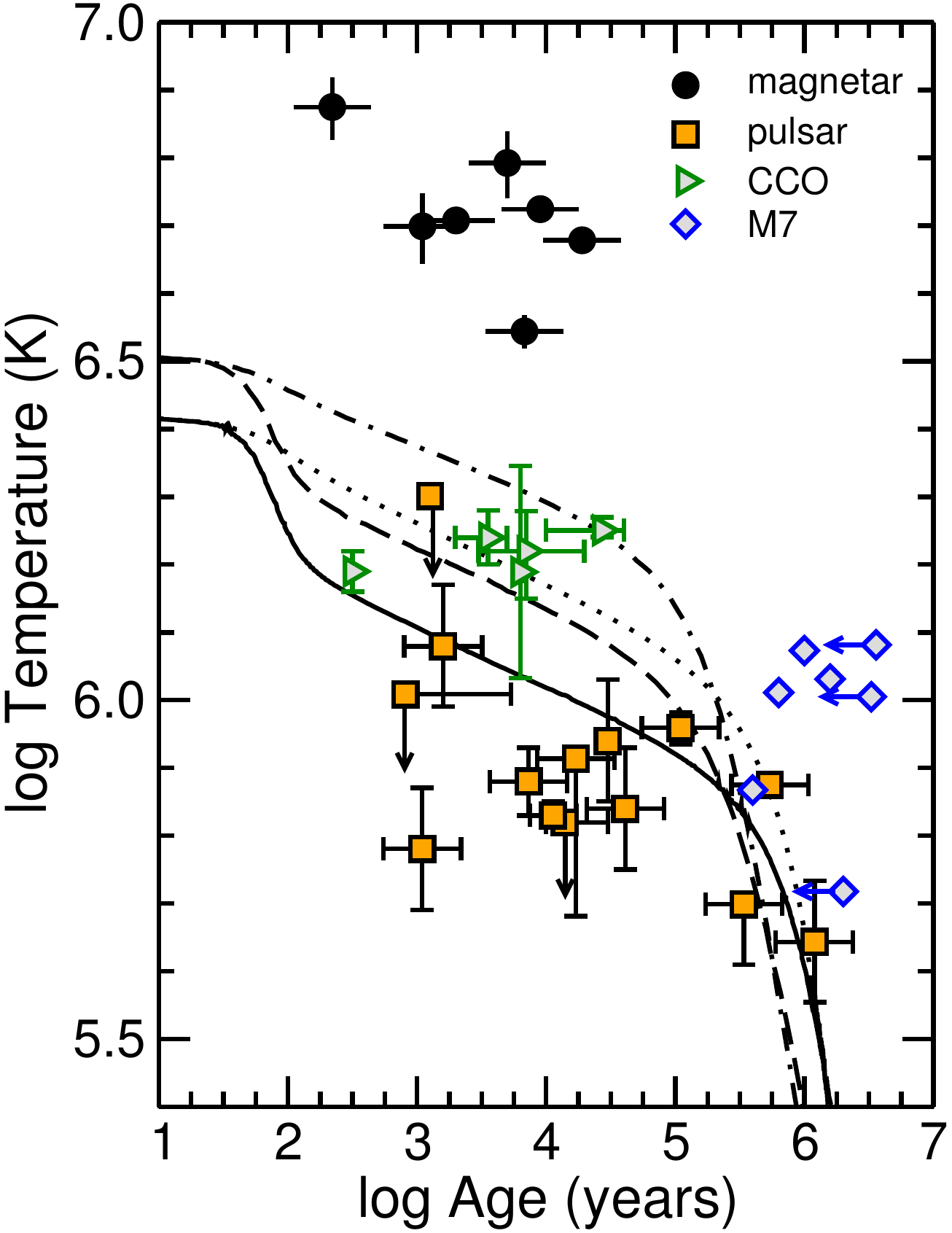}
\caption{\small Cooling-age diagram for INSs with thermal emission, in relation to the cooling scenario with no magnetic field decay (solid, dashed, dotted, and dotted-dahsed lines).
\label{fig_ccurve}}
\end{minipage}
\end{figure}

On the one hand, this is realised by the correlation of a neutron star's thermal luminosity and its inferred magnetic field \cite{pon07}, which holds true for sources with $B\gtrsim10^{13}$\,G (Figure~\ref{fig_BLx}). The standard cooling theory \cite{yak04} falls short at describing the temperature of these INSs (Figure~\ref{fig_ccurve}). Whereas for most neutron stars magneto-rotational and thermal evolution proceed almost independently, the cooling of strongly magnetised sources is affected by field decay and magneto-rotational energy dissipation \cite{agu08,pon09}. 
Evolutionary models that take these effects into account \cite{vig13} show that strong fields at birth can brake the INS spin to an asymptotic value over a relatively short timescale; moreover, field dissipation keeps the stellar crust hot for a longer time than expected from the standard cooling theory. 
While accounting for the spectral and timing properties of individual sources, the model implies an evolutionary link between magnetars and the \msev.
 
At the opposite end of the magnetic field distribution, CCOs challenge the stereotype of a young INS. The very low measured spin-down values of three CCOs (dubbed \textit{anti-magnetars}) imply very low magnetic field strengths and characteristic ages much older than expected from their association with a supernova remnant. Moreover, the lack of detected pulsations in most members indicates that the atmosphere is weakly magnetised \cite{hoh09}. This contradicts the evidence that some sources conceal strong magnetic fields \cite{rea16}. The fact that no old ($>10$\,kyr) anti-magnetar with similar timing properties is recognised in either radio or X-ray surveys \cite{got13b,luo15} implies that their magnetic field is not constant at secular timescales.

CCOs might represent a different outcome of neutron star evolution, where they are recovering from an early episode of field burial by hypercritical accretion \cite{how11}. In this scenario, if a large amount of debris falls back onto the neutron star after the supernova explosion, accretion is likely to occur at such high rates that any field can be hidden \cite{gep99}. This may delay the onset of radio pulsar activity for several $1-100$\,kyr, until the field can diffuse back to the surface \cite{mus95}.
\section{Evolutionary scenarios: issues and open questions}
The role played by the decay of the magnetic field in heating the neutron star crust cannot be overlooked, as it partially explains the observed neutron star diversity. However, even state-of-the-art models of field decay are built over uncertain assumptions, concerning, for instance, the initial field configuration and the level of impurity of the crust \cite{vig13}. At present, the observed radio and X-ray pulsar population is not sufficient to constrain different hypotheses \cite{gul15}. 
While the phenomenology of CCOs has triggered a lively debate over their formation and fate, to our knowledge no attempt has been made in population synthesis to include anti-magnetars as a possible outcome of neutron star evolution. 

The paucity of these INS groups has hindered their use in population syntheses on the Galactic scale. While particular aspects of neutron star evolution and emissivity have been investigated through an extensive number of population syntheses of radio pulsars\footnote{Interestingly, recent results considering only the radio pulsar population show that pulsar evolutionary paths in $P-\dot{P}$ space can signficantly deviate from those expected along lines of constant field, when the decay of the angle between spin and magnetic axes is included in the population synthesis \cite{joh16}.}, overall progress can only be hoped for when the various groups are unified in a `grand scheme' of neutron star evolution \cite{kas10}. There are the works which attempt to join the standard radio pulsar population with those of magnetars and the \msev, through models of field decay \cite{pop10,gul15}. A crucial result is that, to reproduce the number of X-ray thermally emitting INSs we observe to date, the mean strength of the magnetic field distribution at birth has to be significantly higher (and the distribution wider) than one would expect from radio pulsar studies alone. Moreover, the models usually predict a large number of very long spin period neutron stars ($P>20$\,s), which are not observed. 

Also important is to assess the relative contributions of the various INS groups to the total number of neutron stars populating the Milky Way. If each INS group is treated independently, a likely consequence is that the Galactic core-collapse supernova rate cannot account for the entire estimated population of neutron stars \cite{kea08}. The neutron star birthrate derived from population synthesis is even more discrepant from the expected value when field decay is taken into account \cite{gul14,gul15}.
Moreover, while active magnetars are not expected to contribute significantly to the absolute number of observable neutron stars in the Galaxy, the existence of transient sources, which are believed to possess low quiescent X-ray luminosities, also has implications for both the total magnetar birthrate as for the ensemble of the population. The birthrate of CCOs is similarly unclear, either as anti-magnetars (that is, as neutron stars born with intrinsically weak magnetic fields), or within the `hidden magnetic field' scenario. The fraction of neutron stars that undergo a phase of fallback accretion and field submergence is unknown, although these events could commonly take place in the early life of a neutron star (see e.g.~\cite{gep99}, for a discussion). 
\section{A forecast for the \eROS\ survey}
As long as only the X-ray bright end of the radio-quiet INS population is known, the observational and theoretical advances seen in recent years may reach a halt.
The Extended R\"ontgen Survey with an Imaging Telescope Array (\eROS) \citep{pre17} is the primary instrument on the forthcoming Spectrum-RG mission. The four-year \eROS\ all-sky survey (eRASS) is expected to detect a large sample of clusters of galaxies, active galactic nuclei, transient phenomena, and Galactic compact objects, among other interesting case studies. eRASS offers a timely opportunity for a better sampling of a considerable number of neutron stars that cannot be assessed by radio and $\gamma$-ray surveys. Eventually, the identification and investigation of \eROS\ sources at faint X-ray fluxes will help us to test alternative neutron star evolutionary scenarios and constrain the rate of core-collapse supernovae in our Galaxy.

To estimate the number of INSs to be detected by eRASS through their thermal X-ray emission, we performed Monte Carlo simulations of a population synthesis model \cite{pir17}. Briefly, neutron stars are created from a progenitor population of massive stars distributed in the spiral arms of the Galactic disk; after the supernova explosion and the imparted `kick' velocity, their evolution in the Galactic potential is followed while they cool down, isotropically emitting soft X-rays. The expected source count rates and total flux are then computed in the \eROS\ detectors, taking into account the absorbing material in the line-of-sight, the detector and survey properties, and the celestial exposure after four years of all-sky observation. We apply an analytical description of the interstellar medium, based on layers of hydrogen in both atomic and molecular form, and commonly adopted cross-sections. All neutron stars in our simulations cool down at the same fiducial rate.

\begin{figure}[t]
\begin{minipage}{0.5\textwidth}
\includegraphics*[width=\textwidth]{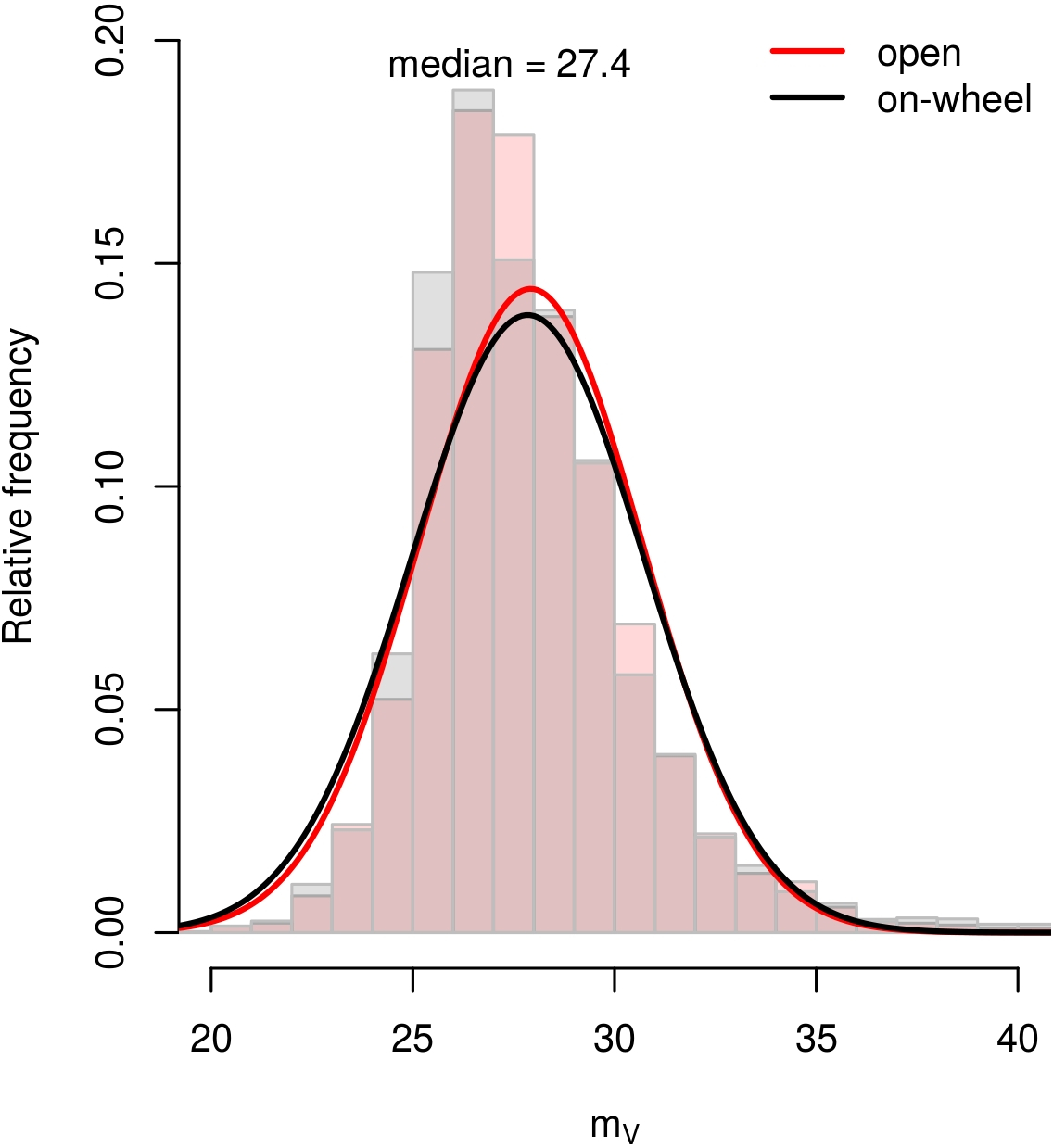}
\end{minipage}
\hspace{0.05\textwidth}
\begin{minipage}{0.4\textwidth}
\caption{\small Histogram of the optical V magnitudes required for ruling out other classes of X-ray emitter among INS candidates, based on X-ray-to-optical flux ratios. We considered in the simulations two possible configurations of optical blocking filters and the resulting total effective area of the seven telescopes, averaged over the field of view \cite{pir17}. A minimum X-ray-to-optical flux ratio of $10^{3.5}$ is assumed for the \eROS-detected sample of neutron stars. Distances to the detected sources are within 400\,pc and 8\,kpc and the accumulated survey exposure ranges from 1.6 to 8\,ks, with a median around 1.9\,ks. The limiting flux is around $10^{-14}$\,erg\,s$^{-1}$\,cm$^{-2}$, which is 50 times deeper than that of the ROSAT Bright Star Catalogue.
\label{fig_histV}}
\end{minipage}
\end{figure}

Our study indicates an expected number of up to $\sim85-95$ thermally emitting INSs\footnote{Depending on the exact configuration of optical blocking filters \cite{pir17}.} to be detected in the eRASS with more than 30 counts ($0.2-2$\,keV).  
In Figure \ref{fig_histV} we show the histogram of V magnitudes required to select promising INS candidates for deep follow-up investigations (assuming a logarithmic X-ray-to-optical flux ratio of $\log(F_X/F_V)= 3.5$, which is sufficient to exclude other classes of X-ray emitters). Although optical follow-up will require very deep observations -- in particular, the identification of the faintest candidates will have to wait for the next generation of extremely large telescopes -- sources at intermediate fluxes can be selected for follow-up investigations using current observing facilities. We anticipate $\sim25$ discoveries in the first years after the completion of the all-sky survey (see \cite{pir17}, for details).

\section{Summary and conclusions}
The observed neutron star phenomenology is arguably governed by the properties of the magnetic field: specifically, by its magnitude at birth, whether it decays or grows during the star's lifetime, and how these characteristics affect the rotational and thermal history of the neutron star. Overall, crucial aspects of neutron star evolution and emissivity cannot be probed through the radio and $\gamma$-ray pulsar population alone, and are not yet described by theory. It is thus important to investigate and survey INSs in X-rays in the hope of discovering missing links in their evolution, and to gain insight into neutron star physics and phenomenology.
The forthcoming all-sky survey of \eROS, expected for launch in September 2018, is a unique oportunity to unveil the X-ray faint end of the neutron star population. Both coordination with photometric and spectroscopic surveys at other wavelengths and catalogue cross-correlation procedures yielding actual probabilities of source identification are required to pinpoint promising INS candidates among the millions of \eROS-detected sources. Dedidated follow-up investigations will tackle the evolutionary state of individual targets to evaluate the impact of alternative scenarios on the evolution and observability of the neutron star population in the Milky Way.
\ack{The collaboration of Axel Schwope and Christian Motch is acknowledged. AMP is financially supported by the Deutsches Zentrum f\"ur Luft- und Raumfahrt (DLR) under grant 50 OR 1511.}
\vfill
\begingroup
\let\clearpage\relax
\include{adsjournalnames}
\bibliographystyle{iopart-num}
\bibliography{pns}
\endgroup
\end{document}

%% file: adsjournalnames.tex
%
%
%
%


\def\ref@jnl{}

\def\aj{\ref@jnl{AJ}}                   
\def\araa{\ref@jnl{ARA\&A}}             
\def\apj{\ref@jnl{ApJ}}                 
\def\apjl{\ref@jnl{ApJ}}                
\def\apjs{\ref@jnl{ApJS}}               
\def\ao{\ref@jnl{Appl.~Opt.}}           
\def\apss{\ref@jnl{Ap\&SS}}             
\def\aap{\ref@jnl{A\&A}}                
\def\aapr{\ref@jnl{A\&A~Rev.}}          
\def\aaps{\ref@jnl{A\&AS}}              
\def\azh{\ref@jnl{AZh}}                 
\def\baas{\ref@jnl{BAAS}}               
\def\jrasc{\ref@jnl{JRASC}}             
\def\memras{\ref@jnl{MmRAS}}            
\def\mnras{\ref@jnl{MNRAS}}             
\def\pra{\ref@jnl{Phys.~Rev.~A}}        
\def\prb{\ref@jnl{Phys.~Rev.~B}}        
\def\prc{\ref@jnl{Phys.~Rev.~C}}        
\def\prd{\ref@jnl{Phys.~Rev.~D}}        
\def\pre{\ref@jnl{Phys.~Rev.~E}}        
\def\prl{\ref@jnl{Phys.~Rev.~Lett.}}    
\def\pasp{\ref@jnl{PASP}}               
\def\pasj{\ref@jnl{PASJ}}               
\def\qjras{\ref@jnl{QJRAS}}             
\def\skytel{\ref@jnl{S\&T}}             
\def\solphys{\ref@jnl{Sol.~Phys.}}      
\def\sovast{\ref@jnl{Soviet~Ast.}}      
\def\ssr{\ref@jnl{Space~Sci.~Rev.}}     
\def\zap{\ref@jnl{ZAp}}                 
\def\nat{\ref@jnl{Nature}}              
\def\iaucirc{\ref@jnl{IAU~Circ.}}       
\def\aplett{\ref@jnl{Astrophys.~Lett.}} 
\def\apspr{\ref@jnl{Astrophys.~Space~Phys.~Res.}}
\def\bain{\ref@jnl{Bull.~Astron.~Inst.~Netherlands}} 
\def\fcp{\ref@jnl{Fund.~Cosmic~Phys.}}  
\def\gca{\ref@jnl{Geochim.~Cosmochim.~Acta}}   
\def\grl{\ref@jnl{Geophys.~Res.~Lett.}} 
\def\jcp{\ref@jnl{J.~Chem.~Phys.}}      
\def\jgr{\ref@jnl{J.~Geophys.~Res.}}    
\def\jqsrt{\ref@jnl{J.~Quant.~Spec.~Radiat.~Transf.}}
\def\memsai{\ref@jnl{Mem.~Soc.~Astron.~Italiana}}
\def\nphysa{\ref@jnl{Nucl.~Phys.~A}}   
\def\physrep{\ref@jnl{Phys.~Rep.}}   
\def\physscr{\ref@jnl{Phys.~Scr}}   
\def\planss{\ref@jnl{Planet.~Space~Sci.}}   
\def\procspie{\ref@jnl{Proc.~SPIE}}   

\let\astap=\aap
\let\apjlett=\apjl
\let\apjsupp=\apjs
\let\applopt=\ao

%% file: pns17_ampires.bbl
\providecommand{\newblock}{}
\begin{thebibliography}{10}
\expandafter\ifx\csname url\endcsname\relax
  \def\url#1{{\tt #1}}\fi
\expandafter\ifx\csname urlprefix\endcsname\relax\def\urlprefix{URL }\fi
\providecommand{\eprint}[2][]{\url{#2}}

\bibitem{agu08}
{Aguilera} D~N, {Pons} J~A and {Miralles} J~A 2008 {\em \apjl\/} {\bf 673}
  L167--L170 (\textit{Preprint} \eprint{arXiv:0712.1353})

\bibitem{gep99}
{Geppert} U, {Page} D and {Zannias} T 1999 {\em \aap\/} {\bf 345} 847--854

\bibitem{got13}
{Gotthelf} E~V, {Halpern} J~P and {Alford} J 2013 {\em \apj\/} {\bf 765} 58
  (\textit{Preprint} \eprint{1301.2717})

\bibitem{got13b}
{Gotthelf} E~V, {Halpern} J~P, {Allen} B and {Knispel} B 2013 {\em \apj\/} {\bf
  773} 141 (\textit{Preprint} \eprint{1307.2146})

\bibitem{gul14}
{Gull{\'o}n} M, {Miralles} J~A, {Vigan{\`o}} D and {Pons} J~A 2014 {\em
  \mnras\/} {\bf 443} 1891--1899 (\textit{Preprint} \eprint{1406.6794})

\bibitem{gul15}
{Gull{\'o}n} M, {Pons} J~A, {Miralles} J~A, {Vigan{\`o}} D, {Rea} N and {Perna}
  R 2015 {\em \mnras\/} {\bf 454} 615--625 (\textit{Preprint}
  \eprint{1507.05452})

\bibitem{hoh09}
{Ho} W~C~G and {Heinke} C~O 2009 {\em \nat\/} {\bf 462} 71--73
  (\textit{Preprint} \eprint{0911.0672})

\bibitem{how11}
{Ho} W~C~G 2011 {\em \mnras\/} {\bf 414} 2567--2575 (\textit{Preprint}
  \eprint{1102.4870})

\bibitem{joh16}
{Johnston} S and {Karastergiou} A 2017 {\em \mnras\/} {\bf 467} 3493--3499
  (\textit{Preprint} \eprint{1702.03616})

\bibitem{kas10}
{Kaspi} V~M 2010 {\em Proceedings of the National Academy of Science\/} {\bf
  107} 7147--7152 (\textit{Preprint} \eprint{1005.0876})

\bibitem{kas17}
{Kaspi} V~M and {Beloborodov} A 2017 {\em ArXiv e-prints\/} (\textit{Preprint}
  \eprint{1703.00068})

\bibitem{kea08}
{Keane} E~F and {Kramer} M 2008 {\em \mnras\/} {\bf 391} 2009--2016
  (\textit{Preprint} \eprint{0810.1512})

\bibitem{luo15}
{Luo} J, {Ng} C~Y, {Ho} W~C~G, {Bogdanov} S, {Kaspi} V~M and {He} C 2015 {\em
  \apj\/} {\bf 808} 130 (\textit{Preprint} \eprint{1507.05381})

\bibitem{man05}
{Manchester} R~N, {Hobbs} G~B, {Teoh} A and {Hobbs} M 2005 {\em \aj\/} {\bf
  129} 1993--2006

\bibitem{mus95}
{Muslimov} A and {Page} D 1995 {\em \apjl\/} {\bf 440} L77--L80

\bibitem{pir17}
{Pires} A~M, {Schwope} A~D and {Motch} C 2017 {\em Astronomische Nachrichten\/}
  {\bf 338} 213--219 (\textit{Preprint} \eprint{1611.07723})

\bibitem{pon07}
{Pons} J~A, {P{\'e}rez-Azor{\'{\i}}n} J~F, {Miralles} J~A and {Miniutti} G 2007
  {\em \apss\/} {\bf 308} 247--257

\bibitem{pon09}
{Pons} J~A, {Miralles} J~A and {Geppert} U 2009 {\em \aap\/} {\bf 496} 207--216
  (\textit{Preprint} \eprint{0812.3018})

\bibitem{pop10}
{Popov} S~B, {Pons} J~A, {Miralles} J~A, {Boldin} P~A and {Posselt} B 2010 {\em
  \mnras\/} {\bf 401} 2675--2686 (\textit{Preprint} \eprint{0910.2190})

\bibitem{pre17}
{Predehl} P 2017 {\em Astronomische Nachrichten\/} {\bf 338} 159--164

\bibitem{rea16}
{Rea} N, {Borghese} A, {Esposito} P, {Coti Zelati} F, {Bachetti} M, {Israel}
  G~L and {De Luca} A 2016 {\em \apjl\/} {\bf 828} L13 (\textit{Preprint}
  \eprint{1607.04107})

\bibitem{tur09}
{Turolla} R 2009 {\em Astrophysics and Space Science Library\/} ({\em
  Astrophysics and Space Science Library\/} vol 357) ed {Becker} W pp 141--163

\bibitem{vig13}
{Vigan{\`o}} D, {Rea} N, {Pons} J~A, {Perna} R, {Aguilera} D~N and {Miralles}
  J~A 2013 {\em \mnras\/} {\bf 434} 123--141 (\textit{Preprint}
  \eprint{1306.2156})

\bibitem{yak04}
{Yakovlev} D~G and {Pethick} C~J 2004 {\em \araa\/} {\bf 42} 169--210
  (\textit{Preprint} \eprint{arXiv:astro-ph/0402143})

\end{thebibliography}
